\documentclass[aps,prl,twocolumn,showpacs,groupedaddress]{revtex4}  % for review and submission 

\usepackage{graphicx}  % needed for figures
\usepackage{dcolumn}   % needed for some tables
\usepackage{bm}        % for math
\usepackage{amssymb}   % for math
\usepackage{epsfig}
\usepackage{subfigure}
\usepackage{latexsym}

\begin{document}
%===================> ADD here your LATEX definitions
\newcommand{\eearrow}   {\mathrm{e^+e^- \rightarrow}}
\newcommand{\righa}     {\rightarrow}
\newcommand{\raa}       {\mbox{$\rightarrow$}}
\newcommand{\m}         {\mathrm}

%=========================================================
% Units
\newcommand{\microns}   {\mbox{$\mu$m}}
\newcommand{\GeVc}      {\mbox{GeV}}
\newcommand{\GeVcc}     {\mbox{GeV}}
\newcommand{\MeVcc}     {\mbox{MeV$/c^2$}}
\newcommand{\MeVc}      {\mbox{MeV$/c$}}
\newcommand{\GeV}       {\mbox{GeV}}
\newcommand{\MeV}       {\mbox{MeV}}
\newcommand{\TeV}       {\mbox{TeV}}
\newcommand{\TeVc}      {\mbox{TeV$/c$}}
\newcommand{\TeVcc}     {\mbox{TeV$/c^2$}}
\newcommand{\invpb}     {\mbox{pb$^{-1}$}}
\newcommand{\invfb}     {\mbox{fb$^{-1}$}}
\newcommand{\pb}        {\mbox{$pb$}}
%=========================================================
\newcommand{\WW}        {\mbox{$W^+W^-$}}
\newcommand{\wboson}    {\mbox{$W^{\pm}$}}
\newcommand{\zboson}    {\mbox{$Z^0$}}
\newcommand{\ee}        {\mbox{$e^+ e^-$}}
\newcommand{\qqg}       {\mbox{$q\bar{q} (\gamma)$}}
\newcommand{\qq}        {\mbox{$q\bar{q}$}}
\newcommand{\pp}        {\mbox{$p\bar{p}$}}
\newcommand{\eeintoqq}  {\mbox{$\eearrow q\bar{q} (\gamma)$}}
\newcommand{\eeintoee}  {\mbox{$\eearrow e^{+} e^{-}$}}
\newcommand{\eeintoZZ}  {\mbox{$\eearrow ZZ$}}
\newcommand{\eeintoll}  {\mbox{$\eearrow l\bar{l}$}}
\newcommand{\eeWWqqqq}  {\mbox{$\eearrow WW \rightarrow q\bar{q} q\bar{q}$}}
\newcommand{\WWqqqq}    {\mbox{$WW \rightarrow q\bar{q} q\bar{q}$}}
\newcommand{\MM}        {\mbox{$\mu^+\mu^-$}}
\newcommand{\zmm}       {{\zboson \raa \MM}}    
%=========================================================
%========================================================================%
\newcommand{\MSB}     {\mbox{$m_{\tilde{b}}$}}
\newcommand{\MNN}     {\mbox{$m_{{\tilde{\chi}}^0_1}$}}
\newcommand{\NN}     {\mbox{$\tilde{\chi}^0_1$}}
\newcommand{\SB}     {\mbox{$\tilde{b}$}}
\newcommand{\sbottom}     {\mbox{$\tilde{b} \rightarrow b \NN$}}
\newcommand{\HZ}     {\mbox{$H \zboson \raa b b \nu \nu$}}
\newcommand{\met}    {\mbox{${\hbox{$E$\kern-0.6em\lower-.1ex\hbox{/}}}_T$}} %missing ET
\newcommand{\mht}    {\mbox{${\hbox{$H$\kern-0.75em\lower-.05ex\hbox{/}}}_T$}} %missing HT
\newcommand{\pt}    {\mbox{$p_T$}}
\newcommand{\ET}    {\mbox{$E_T$}}

\newcommand{\menu} {\mbox{$M_{T}(e,{\ensuremath{\not\!\!\ET                     }})$}}
%========================================================================% 
\newcommand{\Wenujj}    {\mbox{$W(e\nu)+jj$}}
\newcommand{\Wmunujj}   {\mbox{$W(\mu\nu)+jj$}}
\newcommand{\Wtaunujj}  {\mbox{$W(\tau\nu)+jj$}}
\newcommand{\Wtaunuj}   {\mbox{$W(\tau\nu)+j$}}
\newcommand{\Ztautauj}  {\mbox{$Z(\tau\tau)+j$}}
\newcommand{\Znunujj}   {\mbox{$Z(\nu\nu)+jj$}}
\newcommand{\Wtaunubb}  {\mbox{$W(\tau\nu)+b\bar{b}$}}
\newcommand{\Znunubb}   {\mbox{$Z(\nu\nu)+b\bar{b}$}}
\newcommand{\Znunucc}   {\mbox{$Z(\nu\nu)+c\bar{c}$}}
\newcommand{\ttbblnujj} {\mbox{$t\bar{t}\raa b\bar{b} l\nu jj$}}
\newcommand{\ttbblnulnu}{\mbox{$t\bar{t}\raa b\bar{b} l\nu l\nu$}}
\newcommand{\WWincl}    {\mbox{$WW$}}
\newcommand{\ZZincl}    {\mbox{$ZZ$}}
\newcommand{\WZincl}    {\mbox{$WZ$}}
\newcommand{\WZenubb}   {\mbox{$WZ\raa e\nu b\bar{b}$}}
\newcommand{\WZmunubb}  {\mbox{$WZ\raa \mu\nu b\bar{b}$}}
\newcommand{\ZZnunubb}  {\mbox{$ZZ\raa \nu\nu b\bar{b}$}}
\newcommand{\ZZnunucc}  {\mbox{$ZZ\raa \nu\nu c\bar{c}$}}
\newcommand{\toptop}    {\mbox{$t\bar{t}$}}

%========================================================================%

\newcommand{\etmiss}{\ensuremath{\not\!\!\ET                     }}

\def\Missing#1#2{{\mbox{$#1\kern-0.57em\raise0.19ex\hbox{/}_{#2}$}}\ }

\def\vMissing#1#2{\ifmmode
            \vec{#1}\kern-0.57em\raise.19ex\hbox{/}_{#2}
         \else
            {{\mbox{$\vec{#1}\kern-0.57em\raise.19ex\hbox{/}_{#2}$}}\ }
         \fi}

\def\met{\mbox{$\Missing{E}{T}$}}

% The following information is for internal review, please remove them for submission
%\leftline{Version 2.6 as of \today} 
%\leftline{Primary authors:} 
%\leftline{A.-F. Barfuss (anne-fleur.barfuss@ires.in2p3.fr), M.-C. Cousinou (cousinou@cppm.in2p3.fr)}
%\rightline{To be submitted to Phys. Lett. B}
%\rightline{}
%\rightline{}
%\rightline{Comments to {\tt d0-run2eb-015@fnal.gov}}
%\rightline{by xxxx, 2009}

% the following line is for submission, including submission to the arXiv!!
\hspace{5.2in} \mbox{Fermilab-Pub-09/342-E}

\title{Search for pair production of first-generation leptoquarks \\
in {\boldmath$p \bar{p}$} collisions at {\boldmath$\sqrt{s}$} = 1.96 TeV}
% LIST_OF_AUTHORS_R2.TEX                 5/15/09            
%
\author{V.M.~Abazov$^{37}$}
\author{B.~Abbott$^{75}$}
\author{M.~Abolins$^{65}$}
\author{B.S.~Acharya$^{30}$}
\author{M.~Adams$^{51}$}
\author{T.~Adams$^{49}$}
\author{E.~Aguilo$^{6}$}
\author{M.~Ahsan$^{59}$}
\author{G.D.~Alexeev$^{37}$}
\author{G.~Alkhazov$^{41}$}
% ...

\affiliation{\vspace{0.1 in}(The D\O\ Collaboration)\vspace{0.1 in}}
% ...
\affiliation{$^{6}$University of Alberta, Edmonton, Alberta, Canada;
                Simon Fraser University, Burnaby, British Columbia, Canada;
                York University, Toronto, Ontario, Canada and
                McGill University, Montreal, Quebec, Canada}
% ...
\affiliation{$^{30}$Tata Institute of Fundamental Research, Mumbai, India}
% ...
\affiliation{$^{37}$Joint Institute for Nuclear Research, Dubna, Russia}
% ...
\affiliation{$^{41}$Petersburg Nuclear Physics Institute,
                St. Petersburg, Russia}
% ...
\affiliation{$^{49}$Florida State University, Tallahassee, Florida 32306, USA}
% ...
\affiliation{$^{51}$University of Illinois at Chicago,
                Chicago, Illinois 60607, USA}
% ...
\affiliation{$^{59}$Kansas State University, Manhattan, Kansas 66506, USA}
% ...
\affiliation{$^{65}$Michigan State University,
                East Lansing, Michigan 48824, USA}
% ...
\affiliation{$^{75}$University of Oklahoma, Norman, Oklahoma 73019, USA}
% ...
  % input Dzero author list
\date{July 6, 2009}

\begin{abstract}
A search for pair production of first-generation leptoquarks~($LQ$)
is performed with data collected by the D0 experiment
in~$p\bar{p}$ collisions at~\mbox{$\mathrm{\sqrt{s}=1.96~TeV}$} 
at the Fermilab Tevatron Collider. In a sample of data corresponding to $\sim$~1 \invfb\, the search 
has been performed on the final states with two electrons and two jets or one electron, two jets and missing 
transverse energy. We find our data consistent with standard model expectations. The results are combined with 
those found in a previous analysis of events with two jets and missing transverse energy to obtain scalar $LQ$ mass limits.
We set  95\% C.L. lower limits on a scalar $LQ$ mass of 299~\GeVcc, 284~\GeVcc\ and 216~\GeVcc\ for $\beta=1$, 
$\beta=0.5$ and $\beta=0.02$ respectively, where $\beta$ is the $LQ$ branching ratio in the $eq$ channel. This improves the results obtained with a lower luminosity sample 
from Run II of the Tevatron. Lower limits on vector $LQ$ masses with different couplings from 357~\GeVcc\ to  464~\GeVcc\ 
for $\beta=0.5$ are also set using this analysis.

\end{abstract}

\maketitle 
Leptoquarks are conjectured particles, predicted by many extensions \cite{LQsearch} of the standard model (SM).
In such exotic models, transitions between the leptonic and baryonic sectors would be allowed. Thereby, the detection 
of  leptoquarks ($LQ$) could be, among others, the signature of 
compositeness, supersymmetric couplings in R-parity violating models,
Grand Unification models, or technicolor.
Leptoquarks can be scalar or vector fields. It is generally assumed that there is no intergenerational mixing, because it is severely constrained by low-energy experiments, 
and that first-generation $LQ$s couple only to $e$ or $\nu_{e}$ and to $u$ or $d$ quarks.
At the Fermilab Tevatron Collider, pair production of leptoquarks can proceed through
quark-antiquark annihilation (dominant for $M_{LQ} \geq $ 100 \GeVcc) or through gluon fusion, therefore being
independent of the $LQ-e-q$ Yukawa coupling~$\lambda$. Thus the production cross section for scalar leptoquarks
only depends on the strong coupling constant and on the leptoquark mass. In the vector leptoquark case, 
the production 
cross section also depends on the anomalous couplings $\kappa_G$ and $\lambda_G$ of the $LQ$ to the gluon.
At the CERN $e^+e^-$ Collider (LEP), pair production of leptoquarks
could have occurred in $e^{+}e^{-}$ collisions via a virtual  
$\gamma$ or a $Z$ boson in the $s$-channel.
Experiments at the Fermilab Tevatron Collider~\cite{PRDlqp14,CDFRunIIa} and at the LEP Collider~\cite{LEP1} set lower 
limits on the masses of leptoquarks. The H1 and ZEUS experiments at the DESY $e^{\pm} p$ collider HERA
published~\cite{HERA} lower limits on the mass of a first-generation $LQ$ that depend on the coupling $\lambda$.
In the case of single LQ production at LEP or at the Tevatron, the mass limits depend also on $\lambda$~\cite{LEP2}. The branching ratio for  $LQ$ or $\overline{LQ}$ decay into a charged lepton 
and a quark is denoted as $\beta$, so $1-\beta$ is the branching ratio of the reaction $LQ \rightarrow \nu + q$. The branching ratios of 
the three decay modes $LQ \overline{LQ} \rightarrow e e q \bar{q},$ $LQ \overline{LQ} \rightarrow e \nu q \bar{q}$ and 
$LQ \overline{LQ} \rightarrow \nu \nu q \bar{q}$ are then equal to $\beta^{2}$~,  $2  \beta(1-\beta)$ and $(1-\beta)^{2}$, respectively.

In this Letter, we present a search for first-generation leptoquarks for two cases: when both leptoquarks decay to an electron 
and a quark and when one of the leptoquarks decays to an electron and a quark and the other to a neutrino and a quark. The corresponding final states consist of two electrons and two jets ($eejj$) and one electron,  two jets and missing transverse energy ($e \nu jj$).

This study is performed on data collected with the D0 detector~\cite{d0det} in~$p\bar{p}$ collisions at~\mbox{$\mathrm{\sqrt{s}=1.96~TeV}$} 
during Run~II of the Tevatron Collider. 
The D0 detector comprises three main elements. A magnetic central tracking system, which consists of a 
silicon microstrip tracker and a central fiber tracker, is located within a 2 T superconducting solenoidal magnet. 
Three liquid-argon/uranium calorimeters, a central section (CC) covering pseudorapidities~\cite{eta}  $|\eta|$ up to 
$\sim$~1 and two end calorimeters (EC) extending coverage to $|\eta|\simeq 4$, are housed in separate cryostats. 
Scintillators between the CC and EC cryostats provide a sampling of developing showers for $1.1<|\eta|<1.4$. 
A muon system is located outside the calorimeters and covers the region 
$|\eta| < 2$. The luminosity is measured using plastic scintillator arrays placed in front of the EC cryostats.
The data samples for the $e \nu jj$ and $eejj$ analyses are selected with combinations of single electron 
and electron plus jets triggers. The corresponding integrated 
luminosity is $\sim$1 \invfb.

Electrons are defined as clusters of energy deposition in the calorimeters with a high fraction ($> 90 \%$) deposited in the 
electromagnetic (EM) sections. The energy cluster must be isolated from other energy deposits in the calorimeter~\cite{iso} and matched with a charged particle with transverse momentum \pt\ $> 5$~GeV. 
A condition on the value of an electron likelihood  based on a shower shape parameter and conditions on the number of 
tracks in the vicinity of the electron are applied. Electrons that fulfill all the above criteria except the likelihood condition 
are classified as $loose$ electrons. Those which satisfy all criteria are referred to as $tight$ electrons. 

Jets are reconstructed with an iterative cone algorithm~\cite{jet}  with radius 
of 0.5 and a minimal distance $\cal{R} >\ $0.5~\cite{iso} from any EM object. The jet energy scale
(JES) corrections were derived from the transverse momentum
balance in photon-plus-jet events. The missing transverse energy $\met$ is
calculated from all calorimeter cells, and corrected for the jet energy scale and for the transverse momenta of reconstructed muons.

Scalar $LQ$ Monte Carlo samples with masses from 140~\GeVcc\  to 320~\GeVcc\  have been generated with {\sc pythia}~\cite{ref:pythia} using the {\tt CTEQ6L1}~\cite{ref:pdflib,PDFCTEQ61,PDFCTEQ61_bis} 
parton density functions (PDFs). Two processes are generated: $ q \bar{q} \rightarrow  LQ \overline{LQ}$ (dominant for $LQ$ masses 
above 100 \GeVcc) and $gg \rightarrow  LQ \overline{LQ}$~\cite{Hew88}. The $LQ$s are treated as resonances and their isotropic
decay mode is to a $u$ quark and an electron. The {\sc pythia} code has therefore been slightly modified to allow that one of the $LQ$s decays 
into a $d$ quark and a neutrino. The $LQ \rightarrow q l$ vertex depends on the Yukawa coupling $\lambda$ which affects the width of the $LQ$. 
We have taken $\lambda$ equal to the electromagnetic coupling $\sqrt{4  \pi \alpha}$. The next-to-leading order (NLO) cross section of 
scalar $LQ$ pair production has been calculated in Ref. \cite{Kramer}.
To generate the vector leptoquarks, the model described in Ref.~\cite{Belyaev}
and implemented in {\sc comphep}~\cite{comphep} is used.  
In this model, the leading order (LO) cross section depends on the $LQ$ mass and on the anomalous couplings of the $LQ$ to the gluon, $\kappa_G$ and $\lambda_G$.
In the following, three types of couplings have been considered:
``MC" coupling \{$\kappa_G =  1$, $\lambda_G =  0$\},
``YM" coupling \{$\kappa_G =  0$, $\lambda_G =  0$\} and
``MM" coupling \{$\kappa_G = -1$, $\lambda_G = -1$\}. 
We have generated pairs of vector leptoquarks with masses between 200~\GeVcc\ and 480~\GeVcc\  that decay, 
as in the scalar $LQ$ case, into an electron and a quark or into a neutrino and a quark, and we have also 
used a $\lambda = \sqrt{4  \pi \alpha}$.

The main SM backgrounds relevant to these final states are the associated production of jets with 
$Z/\gamma^{*}$ or $W$ boson and top quark pairs in dilepton or semi-leptonic channels. 
Less important contributions come from
$Z/\gamma^{*} \to \tau\tau$ ($\tau\to e$), single top quark decaying into $e$ or $\tau$,
and Diboson final states including jets. Most of the samples were generated with {\sc alpgen}~\cite{ref:alpgen} 
interfaced with {\sc pythia} for parton showering and hadronization. Exceptions are the diboson and single top processes, which were generated with the {\sc pythia} and the {\sc singletop}~\cite{ref:STop} event generators, respectively. The PDFs 
used are {\tt CTEQ6L1}.
The {\sc alpgen} inclusive $W/Z$ production cross section is normalized to the NLO theoretical prediction using K-factors derived by comparing the LO and NLO cross sections in {\sc mcfm}~\cite{ref:mcfm}. All the SM generated backgrounds are normalized to the integrated luminosity of data sample. 

Signal and background Monte Carlo samples are processed through a  {\sc geant}-based~\cite{ref:geant} simulation of the D0 detector and the same reconstruction program as used for the collider data. To model the effects of detector noise and multiple $p\bar{p}$ interactions, each Monte Carlo event is overlaid with a data event from a random $p\bar{p}$ crossing.
Monte Carlo samples pass the same selection criteria as the data samples. But since the efficiency of these selections is different for data and for Monte Carlo, efficiency corrections are applied to the simulated events: the trigger 
probability ($\eta$ and $p_T$-dependent efficiencies for the chosen single electron triggers), a correction for the 
efficiencies of the jet selection, an $\eta$ and $\phi$ dependent correction of the electron selection efficiency, 
and a correction to reproduce the luminosity profile of the data and the distribution along beam axis of the event primary vertex.

In the $eejj$ analysis, 
events are selected with at least two isolated electrons satisfying $tight$ identification criteria, with \pt\ $\geq$ 25~\GeVc\ and at least 
one of the two detected in the central part of the calorimeter ($|\eta | \leq $ 1.1) The selected events must also contain one or more jets with \pt\ $\geq$ 25~\GeVc\ and $|\eta | \leq $ 2.5.
In addition to the main SM backgrounds described above, an instrumental background consists of multijet processes (MJ), and is due to 
the misidentification of jets as electrons. This contribution is extracted from data. A specific 
sample containing events with two ``fake" electrons and at least one additional jet, where a ``fake" electron is an isolated cluster in the 
calorimeter with the usual EM fraction value for a loose electron but shower shape conditions relaxed, is used to reproduce the shapes of the kinematical distributions.
The normalization of the total expected background to the number of data events in two regions of the $M_{ee}$ spectrum ($50 < M_{ee} \leq 80$~\GeVcc\  and $80 < M_{ee} \leq 102$~\GeVcc) gives  the MJ and $Z/\gamma^{*}$ + jets sample contributions. The $t \bar{t}$ and Diboson contributions are normalized to the luminosity. Two normalization factors are extracted and further used to determine the number of background events in 
the sample obtained when the requirement of a second jet with \pt\ $\geq$ 25~\GeVc\ is added.

\begin{figure}[htbp]
\subfigure {\epsfig{figure=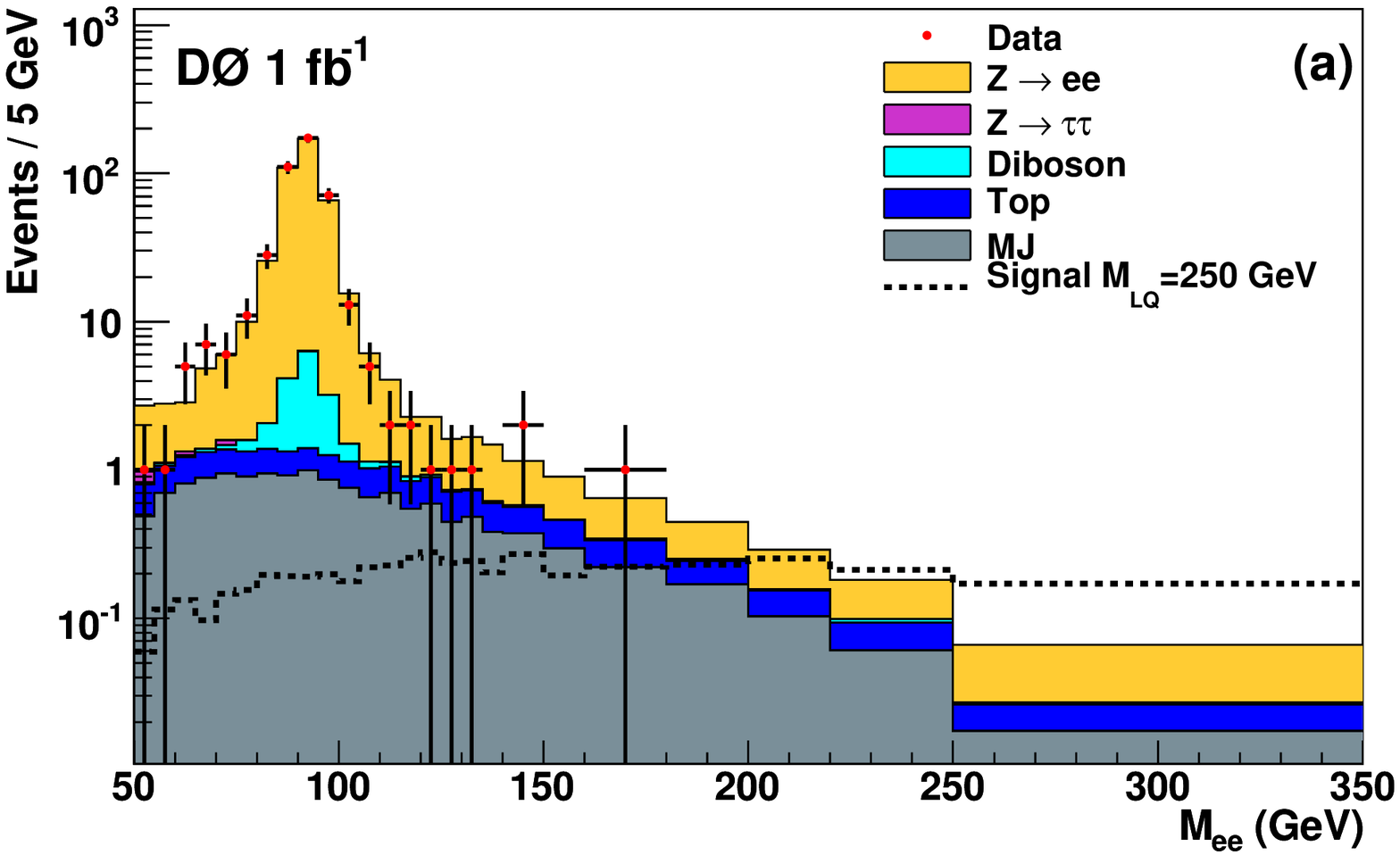,scale=0.42}}
\subfigure {\epsfig{figure=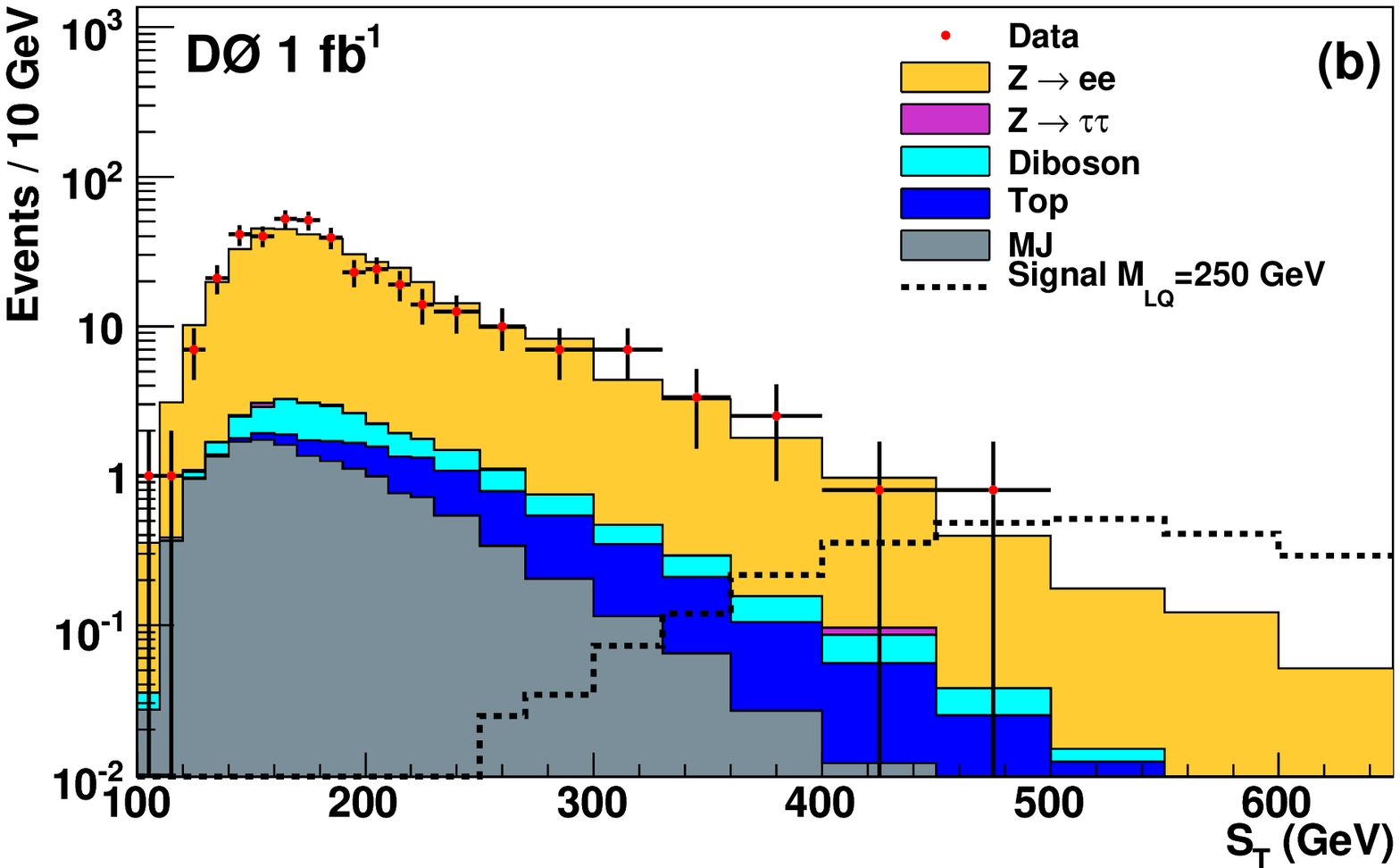,scale=0.42}}
\caption{\label{mee}Distributions of (a) the dielectron invariant mass $M_{ee}$ and (b) $S_T$ for events with $\geq$ 2 jets. The signal 
for a scalar $LQ$ with $M_{LQ}$ = 250 \GeVcc\ has been superimposed.}
\end{figure}

After the requirements of two electrons and two jets, 448 events remain in the data sample, with 449 $\pm$ 13 predicted background events of 
which 91$\%$ originates from the $Z/\gamma^*\to e^+e^-$ samples.
The dielectron invariant mass $M_{ee}$ and the transverse scalar energy $S_T$ (see Fig.~\ref{mee}), defined as the 
scalar sum of the \pt\ of the two electrons and the two highest $E_T$ jets, are used as discriminant variables in 
this analysis. Most $Z/\gamma^*\to e^+e^-$ events are concentrated around the mass of the $Z$ boson ($80<M_{ee}<102$~\GeVcc), 
and the multijet contribution populates the region $S_T<300$~GeV. 
To suppress as much background as possible while minimizing any reduction of signal acceptance, the selections on the $M_{ee}$ and $S_T$ variables have been optimized. For different sets of requirements on these variables, we combine the numbers of expected signal and background events, and their uncertainties, from the bins of the average electron-jet invariant mass distribution to calculate the expected upper limit on the cross section at 95$\%$~C.L. We used a modified frequentist CLs method, based on a likelihood ratio as described in Ref.~\cite{cls}. 
The effects of systematic uncertainties on the signal and background, taking into account correlations, are included in the resulting limits. The best sensitivity is obtained for $M_{ee}>110$~\GeVcc\ and $S_T>400$~\GeV. 
After all selections, no data events remain, for an expected SM background 
of $1.51 \pm 0.12 \mathrm{(stat)} \pm 0.04 \mathrm{(syst)}$ events (see Table \ref{evtsA1}). The acceptance for a scalar $LQ$ with a mass 
between 250 \GeVcc\ and 300 \GeVcc\ varies between $20\%$ and $23\%$. The acceptances for the vector $LQ$s are similar.

\begin{table}[h!]
  \renewcommand{\arraystretch}{1.1}
\centering
\caption{\label{evtsA1}$eejj$ analysis: number of events in each sample after all selections (see text). The two errors on the total expected background correspond to the statistical and systematical uncertainties. The uncertainties quoted on each individual background are only statistical.}
\begin{tabular}{|l|c|}
\hline
 Sample			& Number of events \\
\hline
\hline
Data			&	0	\\
\hline
Total expected background & 1.51 $\pm$ 0.12  $\pm$ 0.04  \\
\hline
 $Z/\gamma^*$ + jets	& 1.11 $\pm$ 0.10 \\
 Multijet		& 0.10 $\pm$ 0.06 \\
 Top	& 0.29 $\pm$ 0.01 \\
 Diboson		& 0.01 $\pm$ 0.01 \\
\hline
\end{tabular}
\end{table}

In the $e \nu jj$ analysis 
we select events containing exactly one isolated electron satisfying $tight$ identification 
criteria with \pt\ $\geq$ 25~\GeVc\ and $|\eta| < 1.1$, and with \met $\geq$ 35 GeV. The selected events must also contain at least 
two high \pt\ jets with $|\eta| < 2.5$, with the leading jet having \pt\ $\geq 40~\GeVc$ and the second leading jet having \pt\ $\geq 25~\GeVc$. 
A veto on a second tight electron with $|\eta| < 2.5$ guarantees that there is no overlap with the $eejj$ analysis.
Multijet processes again contribute to an instrumental background. A fake electron could be present due to misidentification of one jet, 
and the \met could be due to the resolution of the jet energy measurement. Events with $\geq$ 3 jets can thus 
be reconstructed as $e\nu jj$ events. In these events, the \met tends to point in the direction of the fake electron. A triangular 
cut in the $\Delta \phi(e, \met) - \met$  plane is applied ($\Delta \phi(e, \met) \geq \pi - 0.045 \met$ with \met in \GeV) to minimize this background. 

\begin{figure}[htbp]
\includegraphics[scale=0.42]{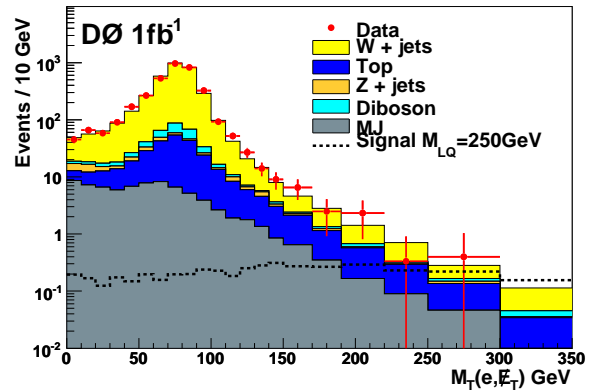}
\caption{\label{menu}Distribution of the variable $\menu$ when the cuts used for the background normalization are applied. 
The signal for a scalar $LQ$ sample with $M_{LQ}$ = 250 \GeVcc\ has been superimposed.}
\end{figure}

In order to model the multijet contribution, a sample containing events with one ``fake" electron and $\geq$ 2 additional jets is created. 
The number of multijet background events is computed using the method described in Ref.~\cite{ttbarMJ}. Two samples of events are used, 
the first one contains events with a $loose$ electron and the second one, which is a subsample of the first one, is composed of events 
with a $tight$ electron. Using the number of events in these two samples
together with the efficiencies for a real and a ``fake" electron to pass the likelihood condition, referred to as $\epsilon_{SM}$ and $\epsilon_{MJ}$ 
respectively, we can determine the number of MJ events.
We measure $\epsilon_{SM}$ as the ratio of the number of Monte Carlo events which pass the likelihood condition over the number of Monte Carlo events 
which fail it and correct for differences between data and simulation. We measure $\epsilon_{MJ}$ directly from data assuming 
that the low $\met$ region ($\met \leq 10$~GeV) is dominated by the multijet background after subtracting a small contribution 
of real electrons determined from Monte Carlo.
The number of Monte Carlo $W$ + jets events is normalized to data within a range of the transverse invariant mass 
of the electron and the \met where the expected number of $LQ$s is very small: $\menu \leq 100$~\GeVcc. 
There is good agreement between data and expected SM background both in number of events and in the shape of the distributions. 
The $\menu$ distribution is shown in Fig.~\ref{menu} with the signal for a scalar $LQ$ sample 
for $M_{LQ}$ = 250 \GeVcc\ superimposed.
The number of data events that pass the selection criteria is equal to 3563 which is in good agreement with the total expected 
background of 3549 $\pm$ 68 events, of which 87$\%$ come from $W$ + jets events. A cut $\menu \geq$ 130~\GeVcc\  strongly reduces this background. Other discriminants are the $\pt$ distributions of the decay 
products of the two $LQ$s. We determined the best $\pt$ cuts as described in the $eejj$ analysis, but using the $S_T$ distribution, where $S_T$ is 
the sum of the \pt\ of the electron, the \pt\  of the two leading jets, and \met. The best expected cross section limits are obtained for a 
cut of 80~\GeVc\ on both the \pt\ of the electron and the $\met$, and the cuts \pt (leading jet) $>$ 40 \GeVc\  and \pt (second jet) $>$~25 \GeVc. 
After all selections, 8 events remain, for an expected SM background of $9.8 \pm 0.8 \mathrm{(stat)} \pm 0.8 \mathrm{(syst)}$ events (see Table \ref{evtsA2}). 
The acceptances are similar for scalar or vector $LQ$s. They range from 18.5$\%$ to 20$\%$ for a $LQ$ mass varying between 250 \GeVcc\ to 300 \GeVcc .

\begin{figure}[h!]
\begin{center}
\subfigure {\epsfig{figure=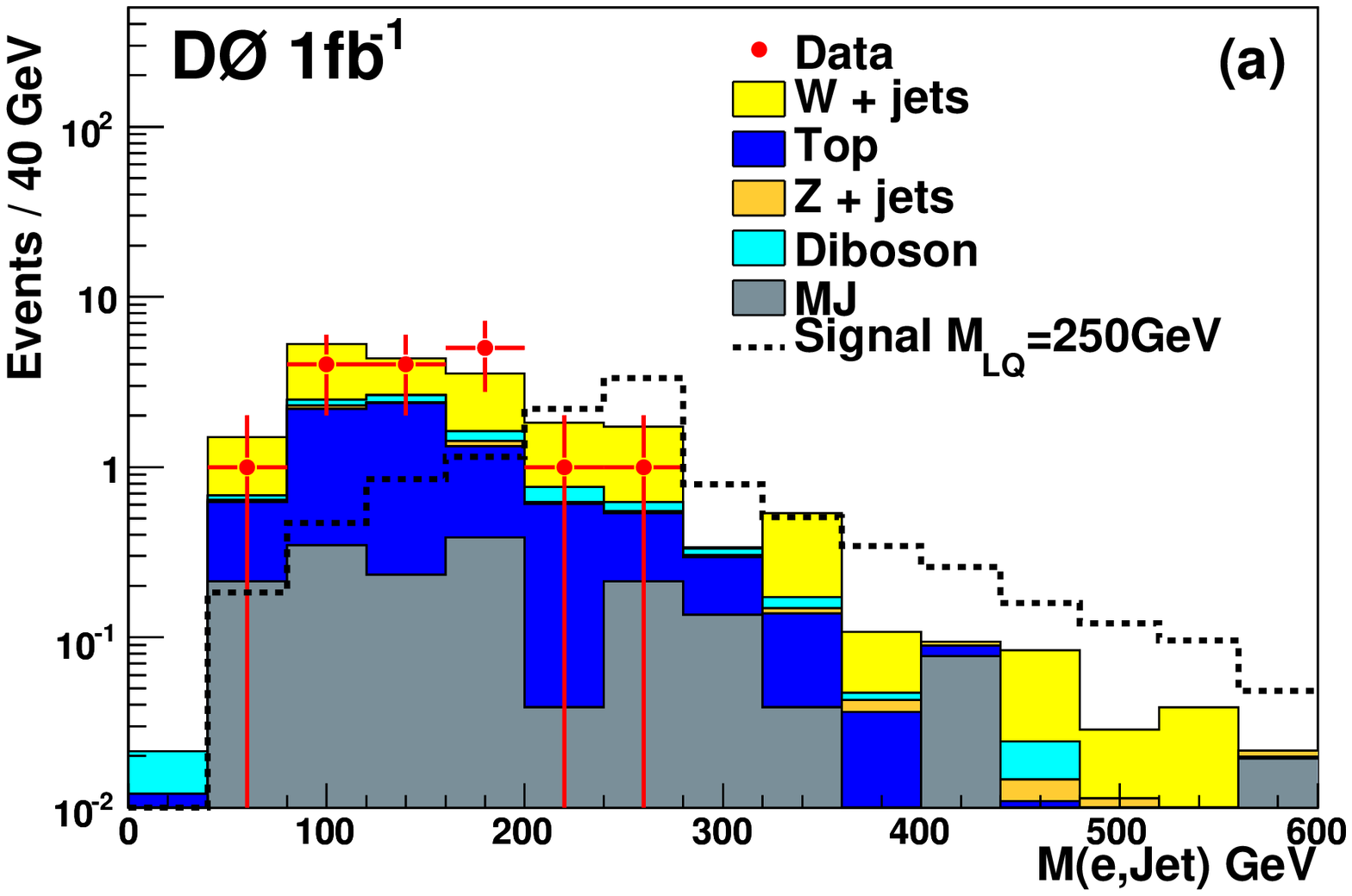,scale=0.42}}
\subfigure {\epsfig{figure=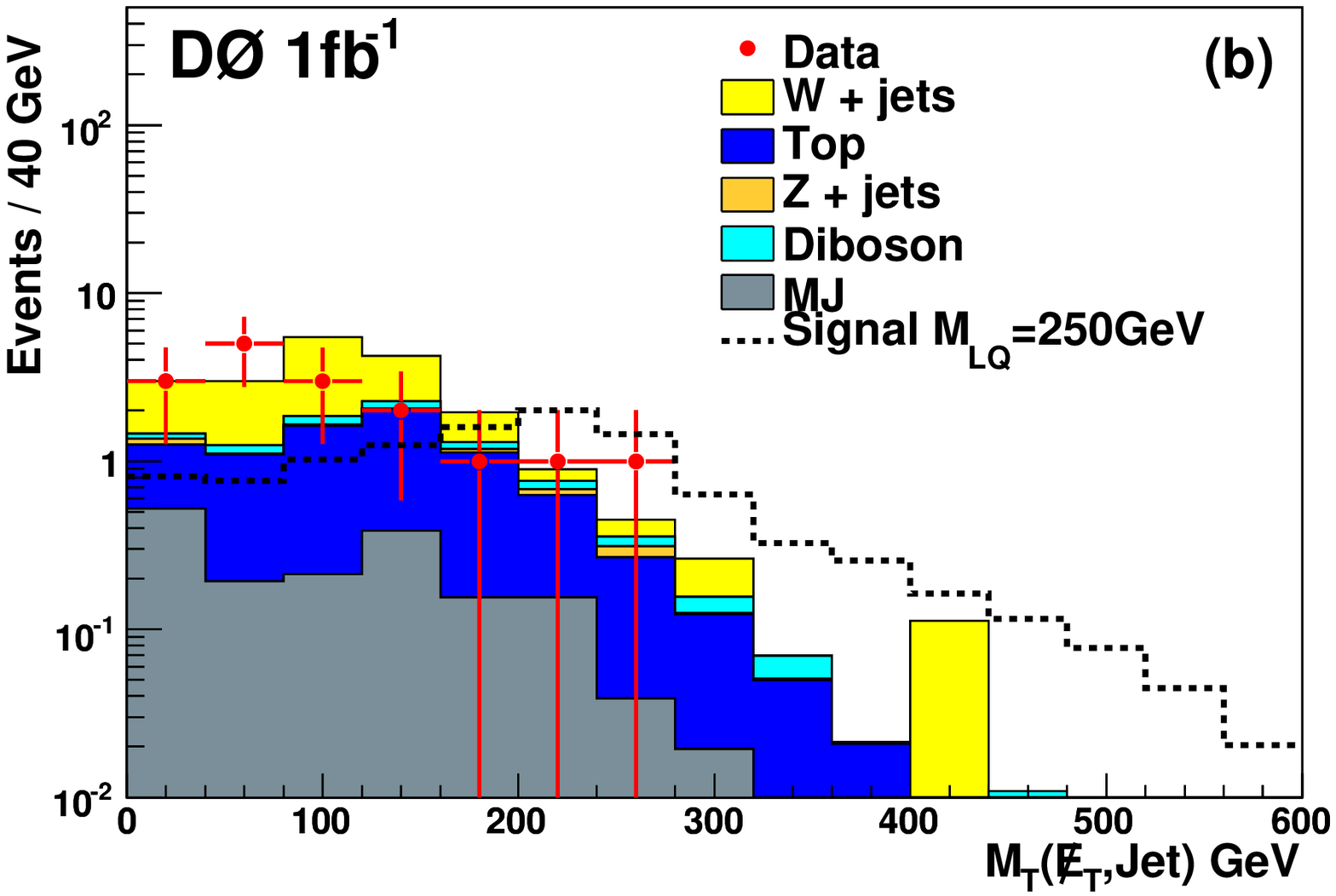,scale=0.42}}
\caption{\label{Mej}Distributions of (a) $M(e, \mathrm{jet})$  and (b) $M_{T}(\met, \mathrm{jet})$ after all cuts. There are two entries 
per event. The signal for a scalar $LQ$ sample with $M_{LQ}$ = 250 \GeVcc\ has been superimposed.}
\end{center}
\end{figure}

In Fig.~\ref{Mej}, the distributions of the masses $M(e,jet)$ and $M_{T}(\met, jet)$ are shown. The signal for a scalar $LQ$ sample for $M_{LQ}$ = 250 \GeVcc\ 
has been superimposed. The agreement is good between data and the SM expectations, both in number of events and in the shape of the distributions. 

\begin{table}[htbp]
  \renewcommand{\arraystretch}{1.1}
\caption{\label{evtsA2}$e \nu jj$ analysis: number of events in each sample after all selections (see text). The two errors correspond to  the statistical and systematic uncertainties.}
\begin{center}
\begin{tabular}{|l|c@{$\,\ \pm\ \,$}c@{$\, \pm$ \,} c|}
\hline
Sample                          & \multicolumn{3}{c|}{\hspace{1mm} Number of events} \\
\hline
\hline
Data                            & \multicolumn{3}{c|}{\hspace{1mm} 8} \\ 
\hline
Total expected background	& 9.8  & 0.8  & 0.8 \\
\hline
$W$ + jets			& 5.0   & 0.7 & 0.3  \\  
Top			& 3.29  & 0.07 & 0.26 \\ 
$Z/\gamma^{*}$ + jets	& 0.15  & 0.06 & 0.01 \\ 
Diboson			& 0.48  & 0.05 & 0.04 \\  
Multijet              & 0.9   & 0.2  & 0.07 \\
\hline
\end{tabular}
\end{center}
\end{table}

The values of the systematic uncertainties are summarized in Table~\ref{Syst}.
Most of them are determined by varying parameters by $\pm 1$ standard deviation. This includes the jet energy scale (JES), the jet energy resolution and the jet identification efficiency (Jet ID). The systematic uncertainty from the correction of the electron identification efficiency (EM ID) is evaluated from the uncertainty on the Monte Carlo/data correction factors and by choosing another parametrization of the correction.
Other systematics uncertainties affect the luminosity, or are computed by measuring the effect of 
the PDF choice on the signal 
acceptances using a different PDF set (20-eigenvector basis {\tt CTEQ6.1M} NLO PDF). The uncertainties due to the propagation into the analyses of uncertainties on the parameters used in the background normalization are referred as background normalization in Table~\ref{Syst}. The SM uncertainties
are the combined relative uncertainties on the expected background due to uncertainties on the cross sections of the SM processes and to different modeling of jet radiation in the $W$ + jets process.
The uncertainties that are shown on the same row are treated as correlated in the determinations of the limits.

\begin{table}[]
\centering
\caption{Summary of systematic uncertainties in $\%$.\label{Syst}}
\begin{tabular}{|l|c|c||c|c|}
\hline
Final state & \multicolumn{2}{c||}{$e e jj$ } & \multicolumn{2}{c|
}{$e \nu jj$ } \\
\hline
Source				&  \hspace{1mm}SM\hspace{1mm}	& \hspace{1mm}Signal\hspace{1mm} & \hspace{1mm}SM\hspace{1mm}	& \hspace{1mm}Signal\hspace{1mm}  \\
\hline
JES			   	&+1.7-2.0& +0.1-0.5  & +1.8-1.3	& +0.9-0.5 \\
Jet resolution		        & 	& & +1.5-0.5 	& \\
Jet ID                          &  0.4		&	0.7 &   0.2    &   0.7 \\
EM ID				& 0.2	&	8	&  1.4 	& 4.2  \\
Luminosity			&  & 6.1 & 2.5  & 6.1  \\
PDF (acceptance)                &     &	5	&	 &	5\\
\hline
Background norm.               & 1.2 & & &  \\
SM uncertainties		& 1.2	& & & 	\\ 
\hline
Background norm.		& &	&  4.4 	& 	\\
SM uncertainties               &  &     & 7.6   &\\
\hline
\end{tabular}
\end{table}

\begin{figure}
\includegraphics[scale=0.42]{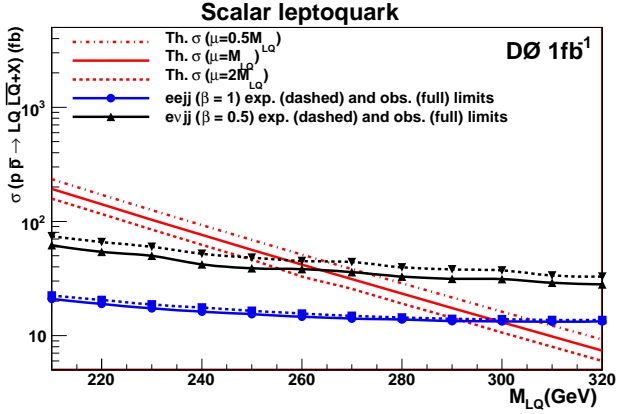}
\caption{\label{limsigmaSLQ}Cross sections as a function of the $LQ$ mass for a scalar leptoquark. The NLO theoretical cross 
sections are drawn for different values of the renormalization and factorization scales: $M_{LQ}$ (solid line), $M_{LQ}/2$ (dot-dashed line) 
and $2M_{LQ}$ (dashed line). 
The horizontal lines correspond to the expected cross section limits (squares and downward triangles) and the observed cross section limits (circles and upward triangles),  both at the 95$\%$ C.L., in 
the $eejj$ channel (blue curves) assuming $\beta =1$ and in the $e\nu jj$ channel (black curves) assuming $\beta=0.5$.}
\end{figure}

No deviations from the SM predictions were observed in our data in either the $eejj$ final state or in the $e \nu jj$ final state and for each individual 
channel we determined cross section limits on the pair production of a first-generation scalar $LQ$ at 95$\%$ C.L. The results are shown 
in Fig.~\ref{limsigmaSLQ} where the expected and observed cross section limits measured in the $eejj$ and $e \nu jj$ final states are displayed as a function of the $LQ$ mass, assuming $\beta=1$ and $\beta=0.5$ respectively. On the same figure the scalar $LQ$ pair production NLO cross sections, 
calculated for different values of the renormalization and factorization scales ($\mu$ = $M_{LQ}$, $M_{LQ}/2$ and $2M_{LQ}$) are also shown.

\begin{figure}
\includegraphics[scale=0.42]{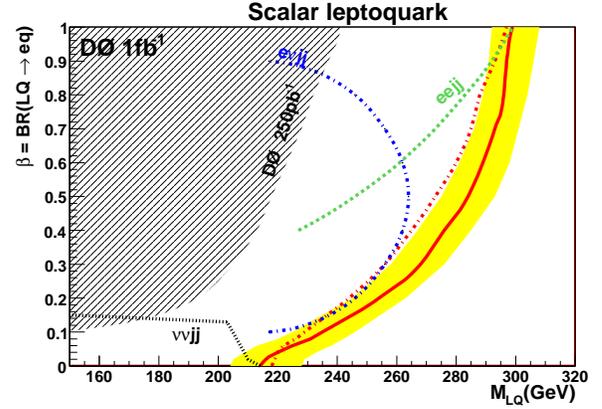}
\caption{\label{massSLQ}Observed (red full line) and expected (red dot-dashed line) mass limits at the 95$\%$ C.L. in the $\beta$ 
versus $LQ$ mass plane for the pair production of first-generation scalar leptoquarks and the nominal signal cross section 
hypothesis ($\mu = M_{LQ}$). The regions to the left of the curves are excluded. The band, around the observed mass limit curve, shows the effect of the theoretical 
uncertainty (see text) on the observed exclusion. The observed limits found individually using each of the three final states are shown for the nominal 
cross section hypothesis ($\mu = M_{LQ}$) and the hatched area is the part of the plane previously excluded by the D0 collaboration 
with a lower luminosity and for the minimal cross section hypothesis ($\mu = 2 M_{LQ}$).}
\end{figure}

In D0 analysis~\cite{nunujj}, using a sample of 2.5 \invfb\ of data with acoplanar jets and missing transverse energy, a search for the 
pair production of first generation scalar leptoquarks both decaying in $\nu q$ has shown no evidence of this production. We combined these three 
analyses to determine expected and observed cross section limits as a function of $\beta$ and $M_{LQ}$. We used the modified frequentist CLs method referenced in the $eejj$ analysis and the JES, PDF and  luminosity systematics uncertainties are 
treated as correlated errors. As an example, the values of the 
observed cross section limits are given in Table~\ref{limLQ} for $\beta=1$ and $\beta=0.5$. For each value of $\beta$, the limit is the LQ
mass value where the experimental cross section limit and the theoretical cross section are equal.
The expected and observed mass limits for factorization and renormalization scales $\mu$ equal to $M_{LQ}$ are summarized in Table~\ref{limmassLQ}. 
They are shown in the $\beta$ - $M_{LQ}$ plane in Fig.~\ref{massSLQ} together with the limits obtained in each final state analysis. The theoretical 
uncertainty on the observed mass limit, reflecting the PDF, normalization and factorization scale uncertainties, is also shown.

\begin{figure}
\includegraphics[scale=0.42]{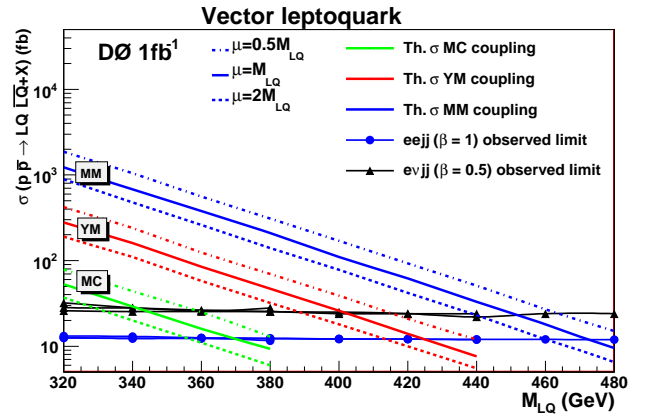}
\caption{\label{limsigmaVLQ}Cross sections as a function of the $LQ$ mass for a vector leptoquark and the three couplings ``MC", ``YM" and ``MM". 
The LO theoretical cross sections are drawn for different values of the renormalization and factorization scales: 
$M_{LQ}$ (solid line), $M_{LQ}/2$ (dot-dashed line) and $2M_{LQ}$ (dashed line). 
The horizontal lines correspond to the observed cross section limits at the 95$\%$ C.L. in 
the $eejj$ channel (circles on blue curves) assuming $\beta$=1 and in the $e\nu jj$ channel (triangles on black curves) assuming that $\beta$=0.5. Small differences in acceptance for different couplings result in marginally different limits shown as the quasi-overlapping curves for each of the channels.}
\end{figure}

\begin{figure}
\includegraphics[scale=0.42]{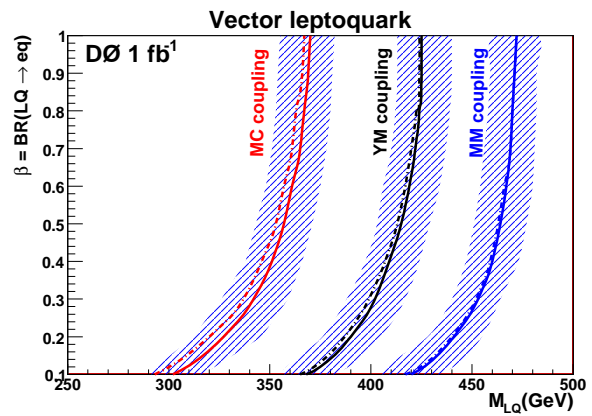}
\caption{\label{massVLQ} Observed (full lines) and expected (dot-dashed lines) mass limits at the 95$\%$ C.L. 
in the $\beta$ versus $LQ$ mass plane for the pair production of first-generation vector leptoquarks. They are shown 
for different couplings (from left to right: the ``MC" coupling, the ``YM" coupling and the ``MM" coupling) and for the nominal 
cross section hypothesis ($\mu = M_{LQ}$). The regions to the left of the curves are exluded. The hatched bands show the effect of the 
theoretical uncertainty (see text) on the observed exclusions.}
\end{figure}

To compute the limit on vector $LQ$ cross sections, as the vector and scalar $LQ$ acceptances are very similar, we use the selections which 
have been found optimal in the search for a scalar $LQ$.
The expected and observed cross section limits for each of the two final states $eejj$ and $e \nu jj$, assuming $\beta=1$ and $\beta=0.5$ 
respectively, are shown in Fig.~\ref{limsigmaVLQ} as a function of the $LQ$ mass. The vector $LQ$ pair production LO cross sections 
are also shown, for each of the three couplings. They are calculated for different values of the renormalization and factorization 
scales $\mu$ = $M_{LQ}$, $M_{LQ}/2$ and $2M_{LQ}$.  We combine these results to get expected and observed cross section limits as a function of $\beta$. 
The values of these limits obtained for $\beta=1$ and $\beta=0.5$ are given in Table~\ref{limLQ}. The expected and observed mass limits for a 
factorization and renormalization scales equal to $M_{LQ}$ are summarized in Table~\ref{limmassLQ}. They are shown in the $\beta$ - $M_{LQ}$ plane 
in Fig.~\ref{massVLQ} for the three couplings. The hatched areas show the effect of the theoretical uncertainties on the observed exclusions.

\begin{table*}[]
\centering
\caption{\label{limLQ}Observed cross section limits (in fb) (95$\%$ C.L.) for a scalar $LQ$ and vector $LQ$ with different couplings as a function of the branching fraction $\beta$.}
\begin{tabular}{|c||c|c|c|c|c|c|c|c|c|c|c|c|c|}
\hline
$M_{LQ}$ (\GeVcc)& 240 & 260 & 280 & 300 & 320 & 340 & 360 & 380 & 400 & 420 & 440 & 460 & 480\\	
\hline
$\beta$		& \multicolumn{13}{c|}{$\sigma$ (fb)   scalar $LQ$  }			\\
\hline
0.5 & 26&24&21&20&19& & & & & & & & \\
1.  & 16&15&14&13&13& & & & & & & & \\
\hline
$\beta$		& \multicolumn{13}{c|}{$\sigma$ (fb)   vector $LQ$ ``MC" coupling }			\\
\hline
0.5 & & &22&21&20&18&17&17& & & & & \\
1.  & & &14&14&13&13&13&12& & & & & \\
\hline
$\beta$		& \multicolumn{13}{c|}{$\sigma$ (fb)   vector $LQ$ ``YM" coupling }			\\
\hline
0.5 & & & &18&18&17&17&17&16&16&15&  &\\
1.  & & & &13&13&12&12&12&12&12&12&  &\\
\hline
$\beta$		& \multicolumn{13}{c|}{$\sigma$ (fb)              vector $LQ$ ``MM" coupling }			\\
\hline
0.5 &  & & & &18&18&17&17&16&16&15&16&16\\
1.  &  & & & &13&12&12&12&12&12&12&12&12\\
\hline
\end{tabular}
\end{table*}

\begin{table}
\centering
\caption{\label{limmassLQ}Expected and observed mass limits (in \GeVcc) for a scalar $LQ$ and vector $LQ$ with different couplings 
as a function of the branching fraction $\beta$, assuming for factorization and renormalization scales $\mu$ = $M_{LQ}$.}
\begin{tabular}{|c||c|c||c|c||c|c||c|c|}
\hline
	&\multicolumn{2}{c||}{scalar $LQ$} &\multicolumn{2}{c||}{``MM" coup.} &\multicolumn{2}{c||}{``YM" coup.} &	\multicolumn{2}{c|}{``MC" coup.} \\
\hline	
  $\beta$ &exp. &obs.&exp. &obs. &exp. &obs. &exp. &obs.\\   
  \hline
0.02& 218&216& & & & & & \\
0.04& 220&220& & & & & & \\
0.06& 222&225& & & & & & \\ 
0.08& 226&231& & & & & & \\  
0.1 & 229&235&417&420&365&368&293&302\\ 
0.2 & 244&254&440&441&387&390&320&327\\
0.3 & 256&268&452&453&399&402&337&342\\
0.4 & 265&276&459&460&407&409&346&351\\
0.5 & 273&284&463&464&413&415&353&357\\
0.6 & 280&289&466&467&417&419&357&361\\
0.7 & 285&293&469&469&420&423&361&365\\ 
0.8 & 288&296&470&470&422&424&363&367\\ 
0.9 & 293&297&471&471&424&425&366&369\\ 
1.0 & 297&299&472&472&424&425&367&370\\
\hline
\end{tabular}
\end{table}

In this analysis of the D0 Run II dataset corresponding to an integrated luminosity of about 1~fb$^{-1}$, we have excluded a 
first-generation scalar $LQ$ with mass varying between 216~\GeVcc\ for $\beta=0.02$ to 299~\GeVcc\ for $\beta=1$ assuming $\mu = M_{LQ}$. 
For  $\mu = 2 M_{LQ}$, the mass limits range from 206~\GeVcc\ to 
292~\GeVcc. These results improve bounds given in previous LQ searches at Tevatron~\cite{PRDlqp14,CDFRunIIa} by $\simeq $ 50~\GeVcc.  We have also excluded vector $LQ$s for different couplings. 
As an example for $\beta=0.5$ and $\mu = M_{LQ}$, lower limits on vector leptoquark masses, varying from 357~\GeVcc\  to 464~\GeVcc, are set for different couplings. These mass limits are the most constraining found in a direct search for first-generation leptoquarks to date.

\vspace{1.0cm}

\begin{acknowledgments}

% acknowledgement_paragraph_r2.tex                         5/15/09
%
We thank the staffs at Fermilab and collaborating institutions, 
and acknowledge support from the 
DOE and NSF (USA);
CEA and CNRS/IN2P3 (France);
FASI, Rosatom and RFBR (Russia);
CNPq, FAPERJ, FAPESP and FUNDUNESP (Brazil);
DAE and DST (India);
Colciencias (Colombia);
CONACyT (Mexico);
KRF and KOSEF (Korea);
CONICET and UBACyT (Argentina);
FOM (The Netherlands);
STFC and the Royal Society (United Kingdom);
MSMT and GACR (Czech Republic);
CRC Program, CFI, NSERC and WestGrid Project (Canada);
BMBF and DFG (Germany);
SFI (Ireland);
The Swedish Research Council (Sweden);
CAS and CNSF (China);
and the
Alexander von Humboldt Foundation (Germany).

\end{acknowledgments}


\begin{thebibliography}{99}
\bibitem{eta} The pseudorapidity $\eta$ is equal to $-\ln[\tan(\theta/2)]$ where $\theta$ is the polar angle relative to the proton beam direction. $\phi$ is the azimuthal angle in the plane transverse to the beam direction.
\bibitem{iso}The fraction of energy in an annular isolation cone of radius $0.2 < \cal{R}$$ <\ 0.4$ must be less than 15\% of the energy in the core cone of radius $\cal{R}$$ < 0.2$, where $\cal{R}$\ $=\sqrt{(\Delta\eta)^2 + (\Delta\phi)^2 }$.
\end{thebibliography}
\end{document}